\documentclass[Journal,letterpaper, BackFigs]{ascelike-new}
\usepackage[utf8]{inputenc}
\usepackage[T1]{fontenc}
\usepackage{lmodern}
\usepackage{graphicx}
\usepackage[figurename=Fig.,labelfont=bf,labelsep=period]{caption}
\usepackage{subcaption}
\usepackage{amsmath}
\usepackage{newtxtext,newtxmath}
\usepackage[colorlinks=true,citecolor=blue,linkcolor=blue, urlcolor=blue]{hyperref}
\usepackage[algonl,boxed]{algorithm2e}
\usepackage{threeparttable}
\usepackage{enumitem}
\usepackage{tabulary}
\DeclareMathOperator*{\argmin}{argmin} 
\setboolean{InsideFigs}{true}\setboolean{BackFigs}{false}

\usepackage{breakcites}
\usepackage{hyperref}

\usepackage[addedmarkup=uline, deletedmarkup=sout, authormarkup=none]{changes}
\definechangesauthor[name=added, color=cyan]{a}
\definechangesauthor[name=deleted, color=red]{d}
\NameTag{\today}
\begin{document}

\title{Performance-Based Risk Assessment for Large-Scale Transportation Networks Using the Transitional Markov Chain Monte Carlo Method}

\author[1]{\href{https://orcid.org/0009-0001-3882-5185}{Anteneh Z. Deriba, S.M.ASCE}}
\author[2,*]{\href{http://orcid.org/0000-0003-0959-6333}{David Y. Yang, Ph.D., A.M.ASCE}}

\affil[1]{Graduate Research Assistant, Dept. of Civil and Environmental Engineering, Portland State University, 1930 SW 4th Avenue, Portland, OR 97201. 
Email: azewdu@pdx.edu}
\affil[2]{Assistant Professor, Dept. of Civil and Environmental Engineering, Portland State University, 1930 SW 4th Avenue, Portland, OR 97201. 
Email: david.yang@pdx.edu}
\affil[*]{Corresponding Author}

\maketitle

\begin{abstract}
 Accurately assessing failure risk due to asset deterioration and/or extreme events is essential for efficient transportation asset management. Traditional risk assessment is conducted for individual assets by either focusing on the economic risk to asset owners or relying on empirical proxies of systemwide consequences. Risk assessment directly based on system performance (e.g., network capacity) is largely limited due to (1) an exponentially increasing number of system states for accurate performance evaluation, (2) potential contribution of system states with low likelihood yet high consequences (i.e., ``gray swan'' events) to system state, and (3) lack of actionable information for asset management from risk assessment results. To address these challenges, this paper introduces a novel approach to performance-based risk assessment for large-scale transportation networks. The new approach is underpinned by the Transitional Markov Chain Monte Carlo (TMCMC) method, a sequential sampling technique originally developed for Bayesian updating. The risk assessment problem is reformulated such that (1) the system risk becomes the normalizing term (i.e., evidence) of a high-dimensional posterior distribution, and (2) the final posterior samples from TMCMC yield risk-based importance measures for different assets. Two types of analytical examples are developed to demonstrate the effectiveness and efficiency of the proposed approach as the number of assets increases and the influence of gray swan events grows. The new approach is further applied in a case study on the Oregon highway network, serving as a real-world example of large-scale transportation networks.
\end{abstract}

\section{Introduction}
Transportation infrastructure systems are crucial to the welfare of a society. Damage to these systems, whether due to deterioration or extreme events, can pose significant risk to the system owner and the traveling public in general. Accurately assessing and effectively mitigating this risk is a major responsibility of transportation agencies worldwide \cite{DHS2013_infra_plan}.

In general, risks of transportation systems are categorized into direct and indirect risks. Direct risk, also known as agency risk, measures the expected economic losses to the asset owners due to adverse events \cite{westernj}. Direct risk can be analyzed for each transportation asset and tallied across all assets within the systems. As a result, it can be effectively managed even for a fairly large system by following a similar practice to portfolio risk management of financial assets \cite{proctorRiskBased2012}. 
Indirect risk of transportation systems, on the other hand, refers to the potential loss of vital functionalities of transportation systems. In practice, it is often assumed that the indirect risk of the entire system can also be isolated at the asset level by, for example, considering the extra travel cost on the asset with damage \cite{bocchiniGeneralized2011,saydamTimedependent2013,fhwaSynchronizing2015}. However, this simplification often ignores the change in route choices of road users and the cascading consequences on other assets within the system.
To overcome this drawback, there has been a growing body of research looking directly into the system functionalities for indirect risk assessment. Several performance indicators are used for this purpose, including connectivity, travel cost, and network flow capacity \cite{changtransportation2010,changBridge2012_,bocchiniConnectivityBased2013,nicholsonFlowbased2016,zhouconnectivity2019,jafinotransport2020,morrisonExploring2022}.

Connectivity between key origin-destination (OD) pairs has been used to measure the loss of accessibility to critical public services, especially during extreme hazard events such as strong earthquakes \cite{zhouResilience2019,dongMeasuring2020,LamImpact2020,morrisonExploring2022}. It is efficient to compute even for large-scale systems, but it becomes less meaningful to less extreme events when connectivity is almost always preserved due to the high redundancy of transportation systems \cite{serdarUrban2022}. 

Alternatively, systemwide travel cost in terms of total travel time and/or distance of all road users has also been leveraged for indirect risk assessment \cite{kiremidjianSeismic2007,shirakiSystem2007,bocchiniStochastic2011,ghosnPerformance2016,vishnuRoad2023}. It considers both the traffic demand from road users and the flow capacity of transportation assets. Although the travel cost is more sensitive to asset damages, risk assessment thereby commonly assumes undisturbed traffic demand after asset damage
(e.g., \citeNP{kiremidjianIssues2007,saydamTimedependent2013,yangRiskInformed2018,capacciProbabilistic2023}), which may not always be valid after extreme events. Moreover, the computation of travel cost can be prohibitively high for risk assessment of large-scale systems.

Network capacity reveals the maximum amount of free-flow traffic that can be accommodated between all or key OD pairs within a transportation system \cite{ahuja1988}. As an intrinsic network characteristic, network capacity has been used to represent the single-commodity freight capacity \cite{nojimaPrioritization1998,morlokMeasuring2004}, an important functionality of transportation infrastructure. Network capacity hinges on the flow capacities of individual assets and their connections in a transportation network. Therefore, it is more sensitive to asset damage than connectivity, while still relatively efficient to compute compared to travel cost. For instance, recent algorithms for determining network capacity have a computational complexity ranging from $O(n^2 \sqrt{m})$ to $O(n^2 m)$, where $n$ and $m$ are numbers of nodes and links respectively \cite{hagbergExploring2008}. Consequently, network capacity is considered a middle ground for indirect risk assessment of large-scale networks and has been widely used \cite{changBridge2012_}.
The present study, therefore, focuses on indirect risk assessment in terms of network capacity.

Irrespective of the specific performance indicators chosen, accurately assessing risk concerning system functionality requires knowing the damage states of all assets. This knowledge is essential for calculating the potential loss in functionality.For instance, consider an OD pair linked by three alternate bridges; it is necessary to know the connectivity of all bridges to check the connectivity of the OD pair. Knowing the damage state of one bridge does not provide sufficient information for assessing connectivity risk between the OD pair. This phenomenon has been referred to in previous studies as the ``network effect'' \cite{yangRiskbased2020,yangRiskbased2022}, which is attributed to the interdependency between assets for realizing system functionality. Because of the network effect, unique challenges arise as the number of assets increases, including the following:
\begin{itemize}
    \item The size of the system states (i.e., the damage states of all assets) grows exponentially with the number of assets. To precisely assess the risk, the need to compute the probabilities and consequences of all system states becomes extremely challenging if not entirely impossible.
    \item The indirect risk may be dominated by those states with low occurrence probabilities but extremely high consequences, usually referred to as gray swan events. Capturing the effects of these gray swan events on indirect risk among the plethora of system states is especially difficult.
\end{itemize}

Previous studies have explored different strategies to address these challenges. To handle the exponentially growing number of system states, most studies rely on Monte Carlo simulation (MCS), which samples only a subset of all system states based on their occurrence probabilities \cite{songMonte2023}.
The assumption underlying this strategy is that the system risk is dominated by more frequent events. This strategy is capable of analyzing large-scale systems, but it cannot effectively consider the influence of potential gray swan events. A large number of samples have to be used in order to sample sufficient gray swan events for risk assessment \cite{soleimaniMultihazard2021,messoreLifecycleCostbasedRisk2021,allenSensitivity2022,luSimulation2023}.
Alternatively, several studies focused on the potential influence of gray swan events \cite{auImportant2003,songSubset2009,echardCombined2013,yangLifecycle2020}. However, such methods do not scale well to large-scale systems due to the so-called curse of dimensionality. Additionally, studies that may scale relatively well \cite{kimSystemReliabilityAnalysis2013} dealt only with specific system performance indicators (i.e., connectivity reliability) and lack versatility for all types of indicators mentioned previously.

In this study, a new approach is proposed to address challenges associated with large number of system states and potential influence of gray swan events in large-scale transportation systems. As a sampling method, the new approach gradually shifts the sampling focus from frequent system states to system states critical to the indirect risk (i.e., gray swan events). The indirect risk is then accurately assessed by analyzing samples throughout the entire process. The proposed approach is adapted from the transitional Markov chain Monte Carlo (TMCMC) method, a sequential sampling method originally developed for Bayesian updating of model parameters. The new approach capitalizes on the tested capacity of TMCMC in high dimensional space to mitigate the challenge caused by the curse of dimensionality. The main contributions of the proposed approach can be summarized as follows:
\begin{itemize}
    \item The approach adapts the problem of indirect risk assessment to the evaluation of the normalization term (also known as evidence) in Bayesian updating.
    \item Due to this adaptation, risk assessment can be carried out with the TMCMC method, which has proven efficiency and effectiveness in high-dimensional space, thus alleviating the curse of dimensionality.
    \item The TMCMC method also gradually shifts the sampling focus from frequent system states to system states critical to the indirect risk. Consequently, the influence of gray swan events can be captured more efficiently. Moreover, the resulting samples give rise to a unique importance measure of networked assets, which can facilitate decision-making.
\end{itemize}

In the following sections, the indirect risk assessment problem is first properly formulated. Based on this formulation, the two aforementioned challenges are further explained before the formal presentation of the new approach. A series of analytical examples are devised to verify the efficiency and effectiveness of the new approach. This is followed by a case study on the Oregon highway network, a real-world example of a large-scale transportation system.

\section{Network Risk Assessment of Transportation Systems}
\subsection{Graph Model for Network Capacity Assessment}
To determine network capacity, a transportation system is modeled as a graph $G = (\mathcal{V}, \mathcal{E})$, where $\mathcal{V}$ is the set of all nodes, usually representing intersections within a transportation system, while $\mathcal{E}$ is the set of links representing road segments between the nodes. The free-flow capacity on each link can be treated as a property of the link. Based on this graph representation, the maximum flow from an origin node $s$ to a destination node $t$ can be formulated as an optimization problem as follows \cite{ahuja1988}:

\begin{equation}
\label{eq:maxflow_formulation}
\begin{aligned}
\text{Maximize:} & \quad v_{st} \\
\text{Subject to:} 
& \sum_{v:(s,v)\in \mathcal{E}} f_{sv} - \sum_{u:(u,s)\in \mathcal{E}} f_{us} = v_{st} \\
& \sum_{v:(t,v)\in \mathcal{E}} f_{tv} - \sum_{u:(u,t)\in \mathcal{E}} f_{ut} = -v_{st} \\
& 0 \leq f_{uv} \leq c_{uv}, \quad \forall (u,v) \in \mathcal{E} \\
& \sum_{u:(u,v)\in \mathcal{E}} f_{uv} = \sum_{u:(v,u)\in \mathcal{E}} f_{vu}, \quad \forall v \in \mathcal{V}\setminus\{s,t\}
\end{aligned}
\end{equation}
where  $v_{st}$ = flow from $s$ to $t$;$f_{uv}$ = flow from node $u$ to node $v$ on the link $(u,v)$ if it exists in $\mathcal{E}$ (otherwise, $f_{uv}$ =0); $c_{uv}$ = capacity of the link $(u,v)$, which is related to the number of lanes and the speed limit on the link; and $\mathcal{V}\setminus\{s,t\}$ = set of all nodes except for $s$ and $t$, respectively. The first and second constraints in Eq. \eqref{eq:maxflow_formulation} indicate that the total flow originating from node $s$ equals the total flow sinking into node $t$. The third constraint is the capacity constraint, i.e., the flow on a link cannot exceed its capacity. The last constraint models the conservation of flows, i.e., the total flow into a node should equal that out of the same node. 

Several algorithms can be used to solve the optimization problem efficiently, even for a large-scale system with thousands to tens of thousands of nodes and links \cite{ahuja1988}. In this study, the preflow-push algorithm is applied based on the implementation in NetworkX, a Python package for analyzing complex networks \cite{hagbergExploring2008}. The network capacity is defined as the sum of the maximum flow values associated with all key OD pairs \cite{nojimaPrioritization1998,changBridge2012_,yangDeep2022}. Therefore, the associated performance indicator can be expressed as follows: 
\begin{linenomath*}
    \begin{equation}
        \label{eq:flow_performance}
        C_{\mathrm{NET}} = \sum_{(s,t) \in \mathcal{K}} v_{st, \mathrm{max}}
    \end{equation}
\end{linenomath*}
where $\mathcal{K}$ = set of key OD pairs; and $v_{st, \mathrm{max}}$ = maximum flow between $s$ and $t$.
\subsection{Network Risk of Large-Scale Transportation Systems}
Damage or failure of transportation assets can reduce or even eliminate the free-flow capacities of the links where the assets are located. Therefore, the indirect risk of a transportation network can be defined as the expected loss in network capacity and expressed as follows \cite{yangDeep2022,yangLifecycle2020}:
\begin{linenomath*}
    \begin{equation}
        \label{eq:indirect_risk}
        \rho_{\mathrm{NET}} = \sum_\mathbf{s} p(\mathbf{s}) \cdot \left| C_{\mathrm{NET},0} - C_{\mathrm{NET}}(\textbf{s)}  \right|
    \end{equation}
\end{linenomath*}
where $\textbf{s}$ is a vector of damages states of all assets; $p(\textbf{s})$ = probability of the system state; and $C_{\mathrm{NET}} (\textbf{s})$ and $C_{\mathrm{NET},0}$ = network capacity given $\textbf{s}$ and that without any damage, respectively.
Due to the network effect, directly applying Eq. \eqref{eq:indirect_risk} is challenging for large-scale transportation systems with a great number of assets. For instance, if each asset has only two states (failure and survival), a system made up of 100 assets amounts to $2^{100} \approx 10^{30}$ system states, whose capacities have to be determined. This computational challenge motivates the prevalent use of risk assessment for individual assets in common transportation asset management systems, e.g., AASHTOWare BrM version 6.4 \cite{AASHTOWare}. Asset-level risk is sometimes quantified by the expected reduction in the asset capacity, and the system risk is approximated by the sum of asset risks. Herein, we refer to the risk thus obtained as additive risk.
Without this approximation, simulation-based methods, most notably MCS, are often the only viable approach to assessing network risk in large systems \cite{songMonte2023}.
It involves randomly sampling system states based on their probabilities and evaluating their consequences. The average consequence of the sampled system states becomes the estimated risk \cite{rokneddinSeismic2014,vishnuRoad2023}.
nventional MCS may struggle to capture rare but significant events, i.e., gray swan events, due to their low occurrence probabilities. This can lead to less accurate risk assessments, especially when system states are dominated by gray swan events.

\begin{figure}
    \centering
    \includegraphics[width=1\textwidth]{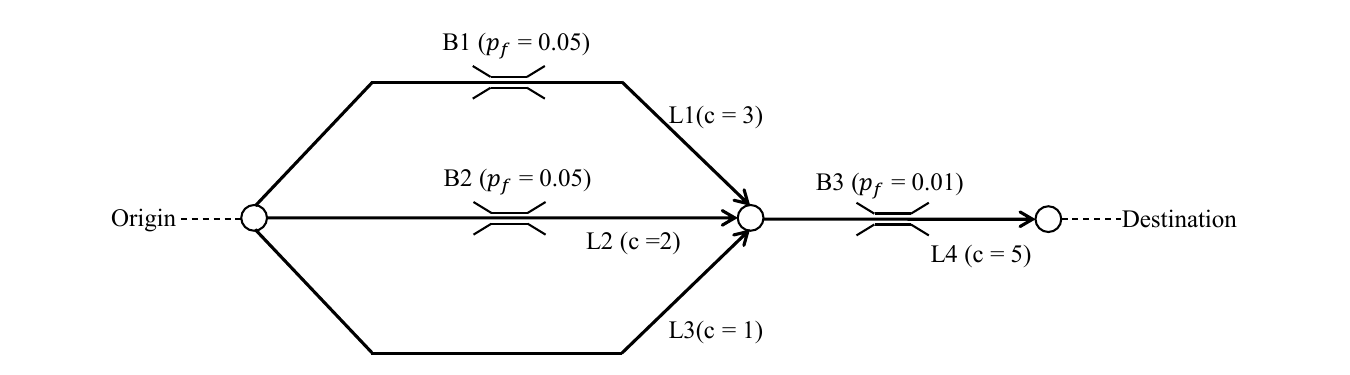}
    \caption{Directed simple graph with three bridges and four links. Probability of failure for bridges and flow capacity for links are shown in brackets; c = capacity and $p_f$ = failure probability.}
    \label{fig:simple_net}
\end{figure}

To best illustrate the drawbacks of existing options, consider the directed simple graph shown in Fig. \ref{fig:simple_net}. The graph represents a simple transportation network made up of four directed links with capacities as shown. For this simple graph, the maximum flow between the origin and the destination is five units, determined directly through inspection. Three bridges with failure probabilities shown in Fig. \ref{fig:simple_net} are located on Link 1, 2, and 4, respectively. These failure probabilities are hypothetical and are used to facilitate the subsequent discussion on risk assessment challenges. They are much higher than those corresponding to the target reliability indices for structural design. However, they can be considered conditional failure probabilities given the occurrence of an extreme event (e.g., a major earthquake).

Based on this network model, the network effect of flow capacity and the gray swan event can be further clarified as follows.
\begin{itemize}
    \item Network effect of capacity drop: Consider the simultaneous failures of Bridges 1 and 3, which would reduce link capacities to zero. The precise failure consequence in terms of drop in network capacity is five. Now consider the additive risk approach. If the consequence of a bridge failure is represented by the loss in the link capacity, the estimated total consequence becomes eight (= 3 + 5). Alternatively, if the reduction in network capacity is analyzed for Bridge 1 and Bridge 3 respectively, the additive indirect consequence is seven (= 2 + 5). This comparison shows that the additive approach is not able to produce accurate estimation of damage consequences when assets are interconnected in a network.
    \item Gray swan event: Each of the three bridges can be in survival or failure states, and enumerating all possible states is trivial because there are only $2^3$ = 8 system states. Precise risk calculation using Eq. \eqref{eq:indirect_risk} gives 4.02\% reduction in network capacity. Now consider simulation approach is used to estimate the risk. Depending on the simulation techniques employed, low probability events with high consequences may not be captured. If 100 samples are used to estimate the risk, the probability of not getting samples with Bridge 3 failure is $0.99^{100}$ (= 0.366). Given no failure of Bridge 3, the conditional system risk becomes 3.05\% by considering only the three failure patterns involving Bridges B1 and B2 and their respective failure probabilities. This shows that failing to account for gray swan events can substantially underestimate the risk.
\end{itemize}
In order to overcome the challenges brought by the network effect and gray swan events, the following section introduces a novel sampling method based on the TMCMC algorithm.
\section{TMCMC for Network Risk Assessment}
\subsection{TMCMC for Bayesian Updating}
Bayesian updating is a statistical inference method used to update prior beliefs of unknown model parameters based on observations of model output \cite{rubinBayesian2015}. Herein, a model can refer to any causal or correlational relationships between input and output. In this context, the prior belief of a model parameter is represented by a prior distribution, and the updated belief is called a posterior distribution. Bayesian updating can result in a more accurate model by refining model parameters. Therefore, it has been widely used within civil engineering for structural system identification, damage assessment, and uncertainty quantification \cite{songAccounting2020,huangStateoftheart2019,simoenDealing2015}. Based on Bayes’ theorem, the posterior joint distribution of model parameters conditioned on observations of model output can be expressed as:
\begin{linenomath*}
\begin{equation}
\label{eq:bayes_theorem}
f(\boldsymbol{\theta}|D) = \frac{f(D|\boldsymbol{\theta}) \cdot f(\boldsymbol{\theta})}{\int_{\boldsymbol{\theta}} f(D|\boldsymbol{\theta}) \cdot f(\boldsymbol{\theta})  d\boldsymbol{\theta}}  
\end{equation}
\end{linenomath*}
where $f(\boldsymbol{\theta})$ = prior probability density function (PDF) of model parameters $\boldsymbol{\theta}$; $f(D|\boldsymbol{\theta})$ = likelihood associated with the observation $D$ of model output, which equals the conditional PDF of $D$ given model parameters $\boldsymbol{\theta}$.
The numerator in Eq. \eqref{eq:bayes_theorem} describes the shape of the posterior PDF. The denominator serves as a normalization constant so that $f(\boldsymbol{\theta}|D)$ satisfies the requirement of a PDF, i.e., the volume under the function should be equal to one. In Bayesian updating, this normalization constant is also called the evidence. Sampling from the posterior PDF is challenging because: (1) the posterior PDF does not always follow an analytical distribution and (2) the evidence is only known by a factor, i.e., only the value of  $f(D|\boldsymbol{\theta}) \cdot f(\boldsymbol{\theta})$ is known or directly computable.

Markov chain Monte Carlo (MCMC) methods are one of the most common Bayesian updating methods \cite{brooksHandbook2011,rubinBayesian2015}. Conventional MCMC methods such as Gibbs sampling and the Metropolis-Hastings algorithm focus on directly sampling from a posterior distribution without knowing the value of the evidence \cite{hastingsMonte1970,rubinBayesian2015}. These methods may suffer from the curse of dimensionality when the number of model parameters increases \cite{beckBayesian2002,robertAccelerating2018}. Several other MCMC algorithms have been proposed to overcome this challenge, e.g., adaptive MCMC \cite{haarioAdaptive2001,haarioDRAM2006,robertsExamples2009, chan2024adaptive}, parallel MCMC \cite{renParallel2007}, and sequential MCMC \cite{cerouSequential2012}.

TMCMC was first proposed for structural dynamics and system identification \cite{chingTransitional2007,betzTransitional2016}. It is one of the sequential methods that generate samples from a sequence of intermediate distributions that gradually converge to the desired posterior distribution. These intermediate samples are generated in so-called stages. Through this process, TMCMC is effective in producing samples of high dimensionality, i.e., large number of model parameters, or with ill-behaving characteristics, e.g., having multiple or sharp peaks \cite{lyeSampling2021}. More importantly for network risk assessment, TMCMC is able to efficiently estimate the evidence term using the intermediate samples from all stages. The fundamentals of TMCMC are presented next, and its adaptation to network risk assessment is described subsequently. More details on the algorithm itself have been given by \citeN{chingTransitional2007} and \citeN{betzTransitional2016}. 

\begin{figure}[h!]
    \centering
    \includegraphics[width=1\textwidth, trim=30 500 30 50, clip]{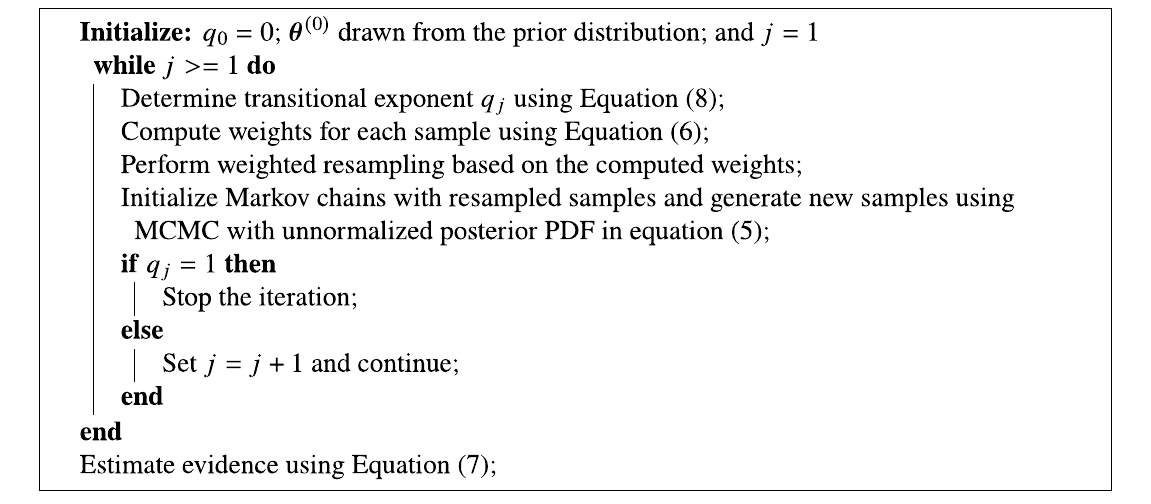}
    \caption{TMCMC algorithm.}
    \label{fig:TMCMC}
\end{figure}

In order to generate samples from the posterior distribution in Eq. \eqref{eq:bayes_theorem}, TMCMC generates intermediate samples in each stage with the following posterior distribution:

\begin{linenomath*}
\begin{equation}
\label{eq:tampering}
f_j (\boldsymbol{\theta}) \propto f \left(D|\boldsymbol{\theta}\right)^{q_j} \cdot f{(\boldsymbol{\theta})}
\end{equation}
\end{linenomath*}
where $j$ = stage number (from 0 to $m$); $f_j(\boldsymbol{\theta})$ = intermediate PDF for stage $j$; $f(\boldsymbol{\theta})$ = prior PDF; and $q_j$ = transitional exponent in stage $j$ to scale the likelihood function. The exponent gradually increases from  $q_0 = 0$ to $q_m = 1$. 
By introducing the transitional exponent, TMCMC starts with samples of model parameters following the prior distribution (due to $q_0=0$). As the exponent gradually increases, samples in intermediate stages are generated with two operations: (1) randomly select samples from the previous stage based on importance weight given in Eq. \eqref{eq:resampling_weights}, and (2) use selected samples as seed to implement MCMC for generating samples following the intermediate distribution in Eq. \eqref{eq:tampering}. The former aims to exploit the existing samples, whereas the latter encourages exploration of new samples. Sample weights in stage $j$ are given as follows:

\begin{linenomath*}
\begin{equation}
\label{eq:resampling_weights}
w^{(j)} = f_{j-1} \left(D|\boldsymbol{\theta}\right)^ {q_j - q_{j-1}}
\end{equation}
\end{linenomath*}
where $f_{j-1}(D|\boldsymbol{\theta})$ = likelihood value that is associated with the sample of the previous stage intermediate distribution; and $q_j$ and $q_{j-1}$ = transitional exponents in the current and previous stages, respectively. In the last stage with $q_m=1$, samples are drawn from the target posterior PDF defined in Eq. \eqref{eq:bayes_theorem}. Moreover, the evidence can be estimated as follows:

\begin{linenomath*}
\begin{equation}
\label{eq:evidence}
S = \prod_{j=1}^m \left[ \frac{1}{N_s} \sum_{k=1}^{N_s} w_k^{(j)} \right]
\end{equation}
\end{linenomath*}
where $w_k^{(j)}$ = weight of sample $k$ at stage $j$; and $N_s$= number of samples. Fig. \ref{fig:TMCMC} illustrates the general procedure of the TMCMC algorithm.

The described process can gradually shift the samples from those following the prior distribution to those following the desired posterior distribution. The overall efficiency of the sequential process hinges on the number of stages. To this end, TMCMC monitors the coefficient of variation (COV) of weights as a new transitional exponent is proposed. Specifically, the transitional exponent in stage $j$ $(j \geq 1)$ is obtained by solving the following optimization problem:

\begin{linenomath*}
\begin{equation}
\label{eq:tamp_exponent}
q_j = \argmin_{q} \left( \left| COV \left\{f_{j-1} \left(D|\boldsymbol{\theta}\right)^{q-q_{j-1}} \right\}_{k=1}^{N_s}  - v_t \right| \right)
\end{equation}
\end{linenomath*}
where $COV\{\cdot\}_{k=1}^{N_s}$ = COV of $N_s$ weights in stage $j$ given a transitional exponent $q$; and $v_t$ = target COV  value. Based on \citeN{chingTransitional2007}, a COV target of 100\% is appropriate to balance the efficiency in intermediate MCMC and the overall efficiency of TMCMC characterized by the number of stages needed.

\subsection{Adaptation of TMCMC for Network Risk Assessment}

\subsubsection{Formulating Network Risk with Multivariate Normal Distributions}

In this project, we focus on the special yet meaningful case of binary assets, namely, each asset has two states: a survival state and a failure state. For the survival state, the link where the asset is located preserves its original capacity, whereas the link capacity decreases (e.g., due to local detours or lane closure) when the asset fails. To fit the network risk assessment in the context of Bayesian updating, a hidden parameter $\xi$ is used to represent the asset state based on the following transformation function:
\begin{linenomath*}
\begin{equation}
\label{eq:tranf_func}
\tau(\xi) = 
\left\{
\begin{array}{ll}
    1 & \text{if } \xi < -\beta \\
    0 & \text{otherwise}
\end{array}
\right.
\end{equation}
\end{linenomath*}
where $\beta$ = reliability index associated with the failure state. For a system with $n$ binary assets, we can similarly define a transformation function $T:\mathbb{R}^n \xrightarrow{}\{0,1\}^n$ as follows:

\begin{linenomath*}
\begin{equation}
\label{eq:tranf_func_n}
T(\boldsymbol{\xi}) = \{\tau (\xi_i)\}
\end{equation}
\end{linenomath*}
where $\boldsymbol{\xi} = \{ \xi_i\}$ for $i = 1,2, \dots , n$. Based on Eq. \eqref{eq:tranf_func_n}, for any system state $\mathbf{s}_k \in \mathcal{S}$, we can find a domain $\Omega_k \subset \mathbb{R}^n$ such that all $\boldsymbol{\xi} \in \Omega_k$ transform to $\mathbf{s}_k$. If $\boldsymbol{\xi}$ follows an independent multivariate normal distribution, the risk associated with a system state $\mathbf{s}_k$ can be expressed as follows:

\begin{linenomath*}
\begin{equation}
\label{eq:state_risk}
L(\mathbf{s}_k)p(\mathbf{s}_k) 
= \int_{\Omega_k} L\left[ T(\boldsymbol{\xi}) \right] \phi(\boldsymbol{\xi})d\boldsymbol{\xi}
\end{equation}
\end{linenomath*}
where $p(\mathbf{s}_k)$ = occurrence probability of system state $\mathbf{s}_k$; $L(\mathbf{s}_k)$ = associated network-level consequence, i.e., $L(\mathbf{s}_k) = C_{\mathrm{NET},0} - C_{\mathrm{NET}}(\mathbf{s}_k)$; and $\phi(\boldsymbol{\xi})$ = PDF of multivariate normal distribution. As a result, by considering all system states, the network risk in Eq. \eqref{eq:indirect_risk} can be expressed in terms of $\boldsymbol{\xi}$ as follows:
\begin{linenomath*}
\begin{equation}
\label{eq:network_risk}
\rho_{\mathrm{NET}}=  \int_{\Omega} L\left[ T(\boldsymbol{\xi}) \right] \phi(\boldsymbol{\xi})d\boldsymbol{\xi}
\end{equation}
\end{linenomath*}
where $\Omega$ = union of $\Omega_k$ for all $k$. The preceding derivation assumes independent asset failures for convenience. This assumption is considered acceptable for several scenarios, such as deterioration-induced failures
\cite{saydamTimedependent2013} or conditional failures given the intensity maps of an extreme event
\cite{guikemaReliabilityEstimationNetworks2009}. In other cases, correlated asset failures may appear and are often quantified by correlation coefficients, e.g., using random field models
\cite{bocchiniStochastic2011,zhangResiliencebasedRiskMitigation2016}. Eq. \eqref{eq:network_risk} can be extended to those cases by replacing $\phi(\boldsymbol{\xi})$ with the PDF of a correlated multivariate normal distribution.
\subsubsection{Estimating Network Risk with TMCMC}

Equating the likelihood function $f(D|\boldsymbol{\theta})$ in Eq. \eqref{eq:bayes_theorem} with $L[T(\boldsymbol{\xi})]$ in Eq. \eqref{eq:network_risk} reveals that Eq. \eqref{eq:network_risk} has the same form as the evidence term in Eq. \eqref{eq:bayes_theorem}. This connection inspires the adaptation of TMCMC for network risk assessment. This equivalence requires non-negative consequence function $L[T(\boldsymbol{\xi})]$, i.e., asset damage does not improve the system performance. This is valid for most performance indicators, including the network capacity used herein.
Accordingly, we can use TMCMC to generate samples of the hidden variable $\boldsymbol{\xi}$ defined in Eq. \eqref{eq:tranf_func_n}. The prior distribution of $\boldsymbol{\xi}$, in this case, is the multivariate standard normal distribution with the same dimensionality as the number of vulnerable assets. In each stage of TMCMC, the sample weights, originally defined in Eq. \eqref{eq:resampling_weights}, can be expressed as follows:

\begin{linenomath*}
\begin{equation}
\label{eq:resampling_weights_2}
w^{(j)} = L\left[T\left(\boldsymbol{\boldsymbol{\xi}}^{(j-1)}\right)\right]^ {q_j - q_{j-1}}
\end{equation}
\end{linenomath*}
where $\boldsymbol{\xi}^{(j-1)}$ = samples in stage $j-1$. Through the sequential process of TMCMC, the estimated network risk can be obtained with Eq. \eqref{eq:network_risk} using the samples from all stages. 

Conceptually, the TMCMC method samples from a posterior distribution function, i.e., samples with greater values of $L\left[ T(\boldsymbol{\xi}) \right] \phi(\boldsymbol{\xi})$ can occur more frequently. In the context of network risk assessment, these frequent samples represent system states with higher contribution to the network risk because these states correspond to larger products between their occurrence probabilities $\phi(\boldsymbol{\xi}$) and the consequences $L[T(\boldsymbol{\xi})]$. Following this logic, if the failure of an asset appears multiple times in the high-risk system states, the asset should be regarded as a high-importance asset in terms of the system performance used for the risk assessment. Therefore, given the last stage samples from TMCMC, denoted as $\mathbf{s}^{(-1)} \in \{0,1\}^{N_s \times n}$ (where $N_s$ = number of samples in each stage and $n$ = number of assets), the importance measure for asset $i$, $\alpha_i$, can be defined as follows:

\begin{linenomath*}
\begin{equation}
\label{eq:asset_impo}
 \alpha_i = \frac{{\sum_{j=1}^{N_s} {s}^{(-1)}_{ji}}}{{N_s}} 
\end{equation}
\end{linenomath*}
where ${s}^{(-1)}_{ji}$ = $j^{th}$ sample for the state of asset $i$. In the formulation of Bayesian updating, this importance measure is equivalent to the posterior failure probability of a specific asset. However, instead of updating the failure probability of an asset, the posterior failure probability inferred from the samples is a risk-informed indicator of asset importance. A greater importance measure may be attributed to a higher prior failure probability, a remarkable loss in system functionality, or both. This asset-level importance measure can assist decision makers with risk-informed ranking of transportation assets.

In this study, the TMCMC method and its adaptation to network risk assessment are implemented based on a preliminary implementation of TMCMC in UQpy, a general-purpose Python toolbox for uncertainty quantification \cite{olivierUQpy2020}. The original code in UQpy was improved and modified to enable parallel computing on high-performance computing facilities.

\section{Method Verification with Analytical Examples}
 
In this section, analytical examples are devised to evaluate the TMCMC method's effectiveness and efficiency in addressing the following two challenges associated with real-world systems: (1) high dimensional risk assessment problem due to an increasing number of vulnerable \mbox{assets (Case I)}, and (2) risk assessment involving network effects and gray swan events (Case II). For both cases, analytical examples are designed to allow for the calculation of precise risk values as benchmarks.

In general, network risk assessment can be carried out using either simulation or non-simulation based methods. The former randomly samples system states based on their occurrence probabilities. The consequences associated with the sampled system states are used to compute the average consequence, i.e., the network risk. Among existing simulation-based methods, crude MCS is the predominant approach due to its ease of implementation and versatility for different system performance indicators \cite{changProbabilisticEarthquakeScenarios2000,shirakiSystem2007,rokneddinSeismic2014,ghoshSeismic2014,ishibashiFramework2020,soleimaniMultihazard2021,messoreLifecycleCostbasedRisk2021,allenSensitivity2022,luSimulation2023}. Non-simulation-based methods attempt to directly and quickly estimate the network risk based on a selective subset of system states \cite{kimAssessmentSeismicRisk2012}. The larger this subset is, the more accurate the approximation becomes. Such methods were originally established for system reliability analysis, i.e., quantification of network disconnection probabilities \cite{kangMatrixbasedSystemReliability2008,songSystemReliabilitySensitivity2009,kangFurtherDevelopmentMatrixbased2012}. \citeN{yangLifecycle2020} further developed the so-called risk-bound method to handle other system performance indicators including network capacity. Based on the previous discussion, the network risk is analyzed using the following three methods: (1) TMCMC method described previously, (2) the crude MCS, and (3) the aforementioned risk-bound method.
To ensure the validity and consistency of the comparison, the following criteria are enforced for the comparison:

\begin{itemize}
    \item For crude MCS, the same number of consequence evaluations as that during the TMCMC implementation is used to stop the simulation.
    \item For the risk-bound method, we use a subset that has the same number of unique system states as that encountered during the TMCMC implementation.
\end{itemize}

The parameters used for TMCMC implementation are presented in Table \ref{table:TMCMC_parameters}. These parameters are either based on the original studies \cite{chingTransitional2007,betzTransitional2016} or assumed in the current study. TMCMC simulation using these parameters are shown to accurately estimate the precise risk. Hyperparameter tuning can be further implemented to determine these parameters. However, this is beyond the scope of the study. Details on the analytical examples and the comparison of results are presented in the following sections.

\begin{table}[htbp]
    \centering
    \caption{TMCMC parameters}
    \label{table:TMCMC_parameters}
    \begin{tabulary}{\linewidth}{LLL} 
        \hline
        \textbf{Parameter} & \textbf{Value} & \textbf{Source}\\ \hline
        Prior standard deviation & 1.0 &  Standard Normal Prior\\ 
        COV threshold for stage evolution & 100\% &\citeN{chingTransitional2007}\\  
        Number of samples per stage & 5000 & Current Study\\ 
        Number of chains during MCMC & 10 & Current Study\\ 
        Number of burn-in samples during MCMC & 1000 & Current Study\\ 
        Number of samples to skip during MCMC & 10 & Current Study\\
        Percentage of resampling & 10\% & Current Study\\ \hline
    \end{tabulary}
\end{table}

\subsection{Case I: Increasing Asset Numbers}
The analytical examples in Case I are designed using the following assumptions:
\begin{itemize}
    \item[(1)] Consider systems with increasing numbers of assets (5, 10, 30, and 50). For each system, the reliability index of an asset is randomly generated following a uniform distribution between zero and three. Then assets were sorted from low to high reliability based on the reliability indices.
    \item[(2)] The failure consequence of an asset (i.e., asset-level consequence) was assumed to be its rank in the sorted list of assets. For instance, the asset with the lowest reliability index (Rank 1 in the sorted list) has a failure consequence of one, whereas the consequence of the most reliable asset is $n$ (where $n$ is the number of assets).
    \item[(3)] To focus on the investigation of asset numbers, the network effect during consequence evaluation was not considered herein. This assumption enables direct summation of the asset consequences to obtain system-level consequences. Specifically, given the system state $\textbf{s}$, the system-level consequence can be expressed as:
    \begin{linenomath*}
    \begin{equation}
    \label{eq:sys_consq}
     C(\textbf{s}) = \sum_{i=1}^{n} c_i \cdot s_i
    \end{equation}
    \end{linenomath*}
    where $s_i$ = state of asset $i$, either 0 or 1 to indicate survival or failure, respectively; $\mathbf{s} = {s_i}$ for $i=1,2,...,n$; and $c_i$ = consequence associated with failure of asset $i$.
    \end{itemize}
Based on the simplification in mentioned in assumption (3), the total system risk $\rho_I$ can be expressed precisely as follows:
    \begin{linenomath*}
    \begin{equation}
    \label{eq:sys_risk}
     \rho_I = \sum_{i=1}^{n} p_{\mathrm{AST},i} \cdot c_i
    \end{equation}
    \end{linenomath*}
    where $p_{\mathrm{AST},i}$  = failure probability of asset $i$.
    
Following the comparison criteria outlined previously, Tables \ref{table: case_I_results_5}--\ref{table: case_I_results_50} present the results of risk estimation using the three methods. The statistics of the estimated risk are derived from 10 different runs to account for the randomness during sampling in TMCMC and MCS methods. These results show that the risk-bound method is accurate and efficient only when the system had a small number of assets (e.g., 5 and 10 assets in this example). In these cases, the risk-bound method considers all or most of the system states that can occur, resulting in accurate risk estimation. As the number of assets increases (e.g., starting with 30 assets), the risk-bound method significantly underestimates the system risk. By contrast, both TMCMC and MCS can accurately estimate the network risk, although TMCMC yields results with more variability. The slightly higher variability of the TMCMC method may stem from the MCMC operation used to seek gray swan events \cite{robertMonteCarloStatistical2005}. Because Case I examples do not involve any gray swan events, such effort only introduces more variability without improving the accuracy in risk assessment.

\begin{table}[htbp]
    \centering
    \caption{Method comparison in Case I with 5 assets (precise risk = 2.205)}
    \label{table: case_I_results_5}
        \begin{tabular}{p{1.4cm}p{1.25cm}p{2cm}p{0.8cm}p{0.8cm}}
            \hline
            \textbf{Method} & \textbf{Average risk} & \textbf{Standard deviation} & \textbf{Max} & \textbf{Min} \\
            \hline
            TMCMC & 2.217 & 0.011 & 2.235 & 2.209 \\
            MCS & 2.202 & 0.003 & 2.205 & 2.198 \\
            Bound & 2.205 & 0.000 & 2.205 & 2.205 \\
            \hline
        \end{tabular}
\end{table}

\begin{table}[htbp]
    \centering
    \caption{Method comparison in Case I with 10 assets (precise risk = 7.527)}
    \label{table: case_I_results_10}
        \begin{tabular}{p{1.4cm}p{1.25cm}p{2cm}p{0.8cm}p{0.8cm}}
            \hline
            \textbf{Method} & \textbf{Average risk} & \textbf{Standard deviation} & \textbf{Max} & \textbf{Min} \\
            \hline
            TMCMC & 7.506 & 0.093 & 7.606 & 7.371 \\
            MCS & 7.531 & 0.009 & 7.540 & 7.520 \\
            Bound & 7.527 & 0.000 & 7.527 & 7.527 \\
            \hline
        \end{tabular}
\end{table}

\begin{table}[htbp]
    \centering
    \caption{Method comparison in Case I with 30 assets (precise risk = 39.45)}
    \label{table: case_I_results_30}
        \begin{tabular}{p{1.4cm}p{1.25cm}p{2cm}p{0.8cm}p{0.8cm}}
            \hline
            \textbf{Method} & \textbf{Average risk} & \textbf{Standard deviation} & \textbf{Max} & \textbf{Min} \\
            \hline
            TMCMC & 39.47 & 0.47 & 39.98 & 38.78 \\
            MCS & 39.45 & 0.08 & 39.54 & 39.34 \\
            Bound & 16.67 & 0.00 & 16.67 & 16.67 \\
            \hline
        \end{tabular}
\end{table}

\begin{table}[htbp]
    \centering
    \caption{Method comparison in Case I with 50 assets (precise risk = 91.41)}
    \label{table: case_I_results_50}
        \begin{tabular}{p{1.4cm}p{1.25cm}p{2cm}p{0.8cm}p{0.8cm}}
            \hline
            \textbf{Method} & \textbf{Average risk} & \textbf{Standard deviation} & \textbf{Max} & \textbf{Min} \\
            \hline
            TMCMC & 91.75 & 0.51 & 92.37 & 90.98 \\
            MCS & 91.35 & 0.12 & 91.45 & 91.18 \\
            Bound & 0.71 & 0.00 & 0.71 & 0.71 \\
            \hline
        \end{tabular}
\end{table}

\subsection{Case II: Risk Involving Network Effects and Gray Swan Events}

To emphasize the network effects and gray swan events, numerical experiments in Case II were designed as follows:
\begin{itemize}
    \item[(1)] Consider a system consisting of multiple assets. Similar to Case I, reliability indices are randomly generated (uniformly between zero and three) and assigned to each asset.
    \item[(2)] To model gray swan events, only a subset of assets with high reliability (hereafter referred to as relevant assets) were considered to induce adverse consequences. In particular, the consequence of an asset is calculated as follows: 
            \begin{linenomath*}
        \begin{equation}
        \label{eq:swan_consq}
        c_i = 
        \left\{
        \begin{array}{ll}
            10^{\frac{(r-1)\beta_i}{n -1}} & \text{if  asset $i$ is relevant} \\
            0 & \text{otherwise}
        \end{array}
        \right.
        \end{equation}
        \end{linenomath*}
        where  $\beta_i$ = reliability index of asset $i$;  $r$ = rank in the ordered asset list; and $n$ = total number of assets.
    \item[(3)] To introduce network effects, the simulation adheres to two rules. First, regardless of the number of failed assets, the system-level consequence is incurred only if at least one of the relevant assets fails. Second, in cases where more than one relevant asset fails, the network consequence is determined by the product of individual failure consequences. Therefore, given a system state and a set of relevant assets $\mathcal{U}$, the system-level consequence can be expressed as:
    \begin{linenomath*}
        \begin{equation}
        \label{eq:sys_consq_swan}
         C(\textbf{s}) = \prod_{i\in \mathcal{U}} c_i^{s_i}
        \end{equation}
    \end{linenomath*}
        where $s_i$ = state of asset $i$; and $c_i$ = consequence defined in Eq. \eqref{eq:swan_consq}.
    \end{itemize}

If only a small subset of assets (among all assets of a system) is relevant to system risk, the precise risk can be determined by enumerating the probabilities and the consequences of all system states involving the failures of relevant assets. Therefore, the precise system risk can be expressed as:
    \begin{linenomath*}
    \begin{equation}
    \label{eq:sys_risk_swan}
     \rho_{II} = \sum_{k \in \mathcal{Q}} p_{f}(\textbf{s}_k) \cdot  C(\textbf{s}_k)
    \end{equation}
    \end{linenomath*}

where $p_{f}(\textbf{s}_k)$ and $C(\textbf{s}_k)$ = failure probability and consequence, respectively, associated with system state $k$; and $\mathcal{Q}$ = set of system states involving relevant assets. For instance, a system with five relevant assets (among all assets) has 31 (= $2^5-1$) relevant system states with consequences.

Due to the design for system-level consequences [Eqs. \eqref{eq:swan_consq} and \eqref{eq:sys_consq_swan}] and the exponential approximation for standard normal CDF tails \cite{temme2024}, all system states involving relevant assets will have similar contributions to the network risk. Hence, Case II examples represent an extreme case involving gray swan events because all system states involving relevant assets influence the system risk equally, regardless how unlikely a state is to occur. Therefore, Case I and Case II examples form a theoretical bound for how real-world networks may behave.

\begin{figure}
    \centering
    \includegraphics[scale=0.75, trim={0.8cm 8cm 0 2.5cm}, clip]{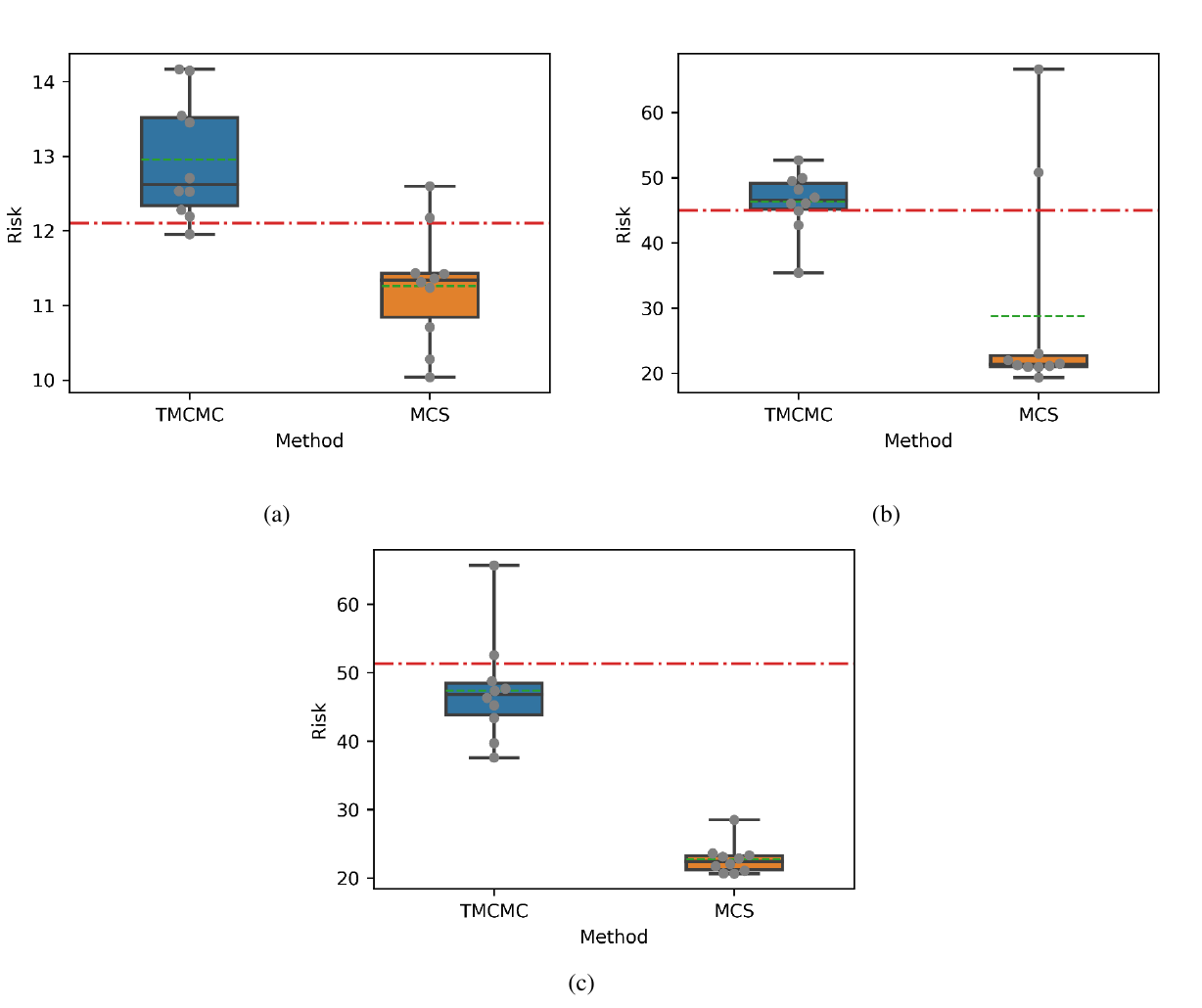}
    \caption{Risk assessment results for Case II examples with varying number of relevant assets: (a) 3 relevant assets out of a total of 30 assets; (b) 5 relevant assets out of a total of 30 assets; and (c) 5 relevant assets out of a total of 50 assets.}
    \label{fig:box-whisker_swan}
\end{figure}

Similar to Case I, each method was implemented 10 times to derive statistics of the estimated risk. Three scenarios were analyzed including three and five relevant assets out of a total of 30 assets (denoted 3/30 and 5/30 scenarios, respectively) as well as five relevant assets out of a total of 50 assets (5/50 scenario). Fig. \ref{fig:box-whisker_swan} shows the estimated risks using both TMCMC and MCS methods in the form of box-whisker plots. The box encloses data from the 25th to the 75th percentiles, and the whiskers are minimum and maximum values. The solid and the dash lines between the whiskers are median and mean values, respectively.

Results indicate that the TMCMC method is able to accurately estimate the precise network risk in Case II examples. 
In particular, Figs. \ref{fig:box-whisker_swan}(a and b) illustrate the effect of different amounts of gray swan events. The 3/30 scenario shown in Fig. \ref{fig:box-whisker_swan}(a) included only 7 relevant system states, and the 5/30 scenario involved 31 relevant system states. Fig. \ref{fig:box-whisker_swan}(a) shows that both MCS and TMCMC closely estimate the precise risk value of 12.107. The mean values from the 10 runs were 12.896 for TMCMC and 11.257 for MCS, respectively. However, as more gray swan events were involved (as in 5/30 scenario), the performance gap between the two methods became more prominent. Namely, the TMCMC method can accurately estimate the precise risk, and the MCS method frequently underestimated the risk and also yielded an impractical amount of variability among different runs.
Comparing Fig. \ref{fig:box-whisker_swan}(b and c) reveals the effect of difficulty in finding gray swan events. Although 5/30 and 5/50 scenarios both have 31 relevant system states, it is more difficult for the MCS method to sample them among the $2^{50}$ system states in the 5/50 scenario. This is reflected in Fig. \ref{fig:box-whisker_swan}(c) where all 10 runs of MCS in the 5/50 scenario yield significant underestimates of the system risk. In practice, situations similar to Fig.  \ref{fig:box-whisker_swan}(c) can be more misleading than Fig. \ref{fig:box-whisker_swan}(b) because an analyst may consider that the system risk is converged across multiple MCS runs and thus falsely conclude that the MCS result is accurate. Overall, Fig. \ref{fig:box-whisker_swan} indicates that the MCS method lacks reliability in the estimated risk, particularly when gray swan events are involved.

\section{Case Study on Oregon Highway Network}

The Oregon highway network comprises thousands of links, many of which include one or more vulnerable bridges. These links with bridges are referred to hereafter as bridge links. The Oregon highway network serves as a typical example of a large-scale transportation system. Each bridge link is treated as a single asset during risk assessment. Similar to the previous analytical examples, each asset is assumed to have a reliability index randomly selected from a uniform distribution ranging from zero to three. In practice, one can derive the reliability index of a bridge link using reliability analysis methods for a series system \cite{ditlevsenStructural1996,derkiureghianStructural2022} based on the reliability indices of all bridges on the link. However, system reliability analysis is beyond the scope of this study and is not implemented here. Because the reliability indices are randomly assigned, the results of this case study are solely used for algorithmic validation and should not be considered the true risk associated with the Oregon highway network. 

\subsection{Developing Graph Model for Oregon Highway Network}

In general, transportation systems are best represented as directed multigraphs. OSMnx, a Python package to extract geospatial features from OpenStreetMap (OSM), can generate these multigraphs based on geospatial data \cite{boeingOSMnx2017}. However, determining network capacity for directed multigraphs is challenging due to flow distribution ambiguities among different links connecting two nodes \cite{newmanNetworks2018}. Therefore, a directed simple graph was created using OSMnx, where only the shortest link among all links connecting the two nodes was retained. 

For this case study, the graph model for the Oregon highway network is constructed using roadway segments labeled as motorways, trunk, and primary in OSM. These segments include the majority of bridges managed by Oregon Department of Transportation (ODOT).
\begin{figure}
    \centering
    \includegraphics[scale=1, trim={0cm 2cm 0 1cm}, clip]{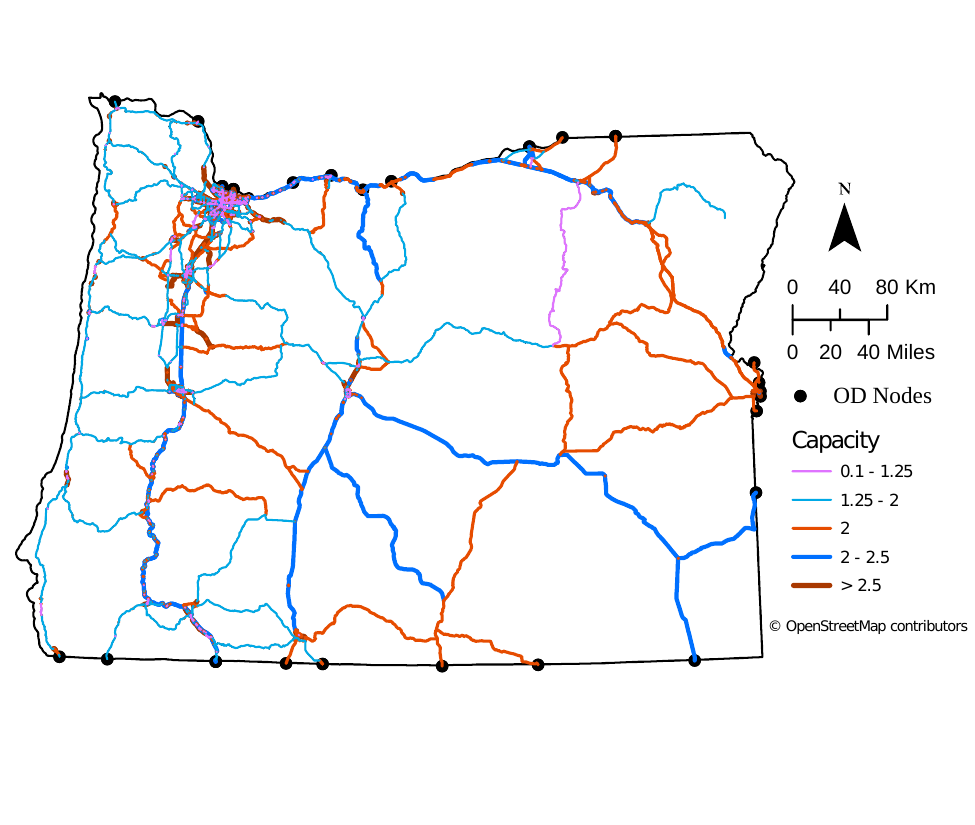}
    \caption{Oregon highway network map (Map data $\copyright$ OpenStreetMap contributors.)}
    \label{fig:Oregon_HW}
\end{figure}
Fig. \ref{fig:Oregon_HW} displays the highway network under consideration. There are 31 boundary nodes representing entrances into and/or exits from Oregon. From these boundary nodes, OD pairs were selected if two boundary nodes are at least 50km apart. In total, 94 OD pairs were considered to determine the network capacity of the Oregon highway network. In practice, this indicator corresponds to the network throughput. The flow capacity of each link is estimated using the following equation:
\begin{linenomath*}
\begin{equation}
\label{eq:nom_flow}
 c_e = n_{\mathrm{lane}} \cdot \frac{v_e}{v_{\mathrm{hwy}}}
\end{equation}
\end{linenomath*}
where $c_e$ = capacity of link $e$; $n_{\mathrm{lane}}$ = number of lanes on the link; $v_e$ = speed limit (mi/h); $v_{\mathrm{hwy}}$ = nominal speed on a highway link, assumed to be 60 mi/h (96.56 km/h) in this study. The lane numbers and the speed limits of all links are available from the OSM geo-database. If a link contained multiple lane numbers and speed limits, the minimum lane number and the average speed limit were used with Eq. \eqref{eq:nom_flow} to compute the link capacity.

The graph depicted in Fig. \ref{fig:Oregon_HW} comprises 6,437 nodes and 10,637 links. Among these links, 1,938 are vulnerable bridge links (i.e. assets). In the event of failure, all traffic on the link is assumed to be diverted to a detour on local roads. Because these local roads are not explicitly modeled in the graph, detours were symbolized by reducing the capacity of a failed bridge link to 0.333 (equivalent to one lane at 20 mi/h or 32.19 km/h) while retaining the link in the graph. This detour-induced reduction in link capacity consequently diminishes the network throughput. Normalizing this reduction by the throughput of the intact network defines the indirect risk associated with bridge failures.

\subsection{Indirect Risk Assessment for Oregon Highway Network}

The indirect risk assessment of this network involves 1,938 binary random variables representing survival or failure of bridge links. As demonstrated by the results in earlier analytical examples, nonsimulation methods like the risk-bound approach are not suitable for handling such a large-scale problem. Moreover, it is unknown whether gray swan events significantly contribute to the network risk. To address these challenges, the TMCMC method proposed herein was employed for network risk assessment. The parameters utilized for this assessment were similar to those used in the analytical examples (Table \ref{table:TMCMC_parameters}), with two adjustments: increasing the number of samples per stage from 5,000 to 8,000 and raising the number of chains from 10 to 80. These adjustments were necessary to accommodate the increased scale of the risk assessment problem.

The proposed method successfully sampled 44,123 unique system states among approximately $2.49 \times 10^{583}$ possible system states. The resulting estimated network risk was 0.3262, indicating an expected 32.62\% reduction in throughput based on the failure probabilities of bridge links. The computation took 23.41 hours on a stack server equipped with 4 Intel Xeon Gold 6230 CPUs (2.10 GHz base clock speed) and a total of 160 logical processors. 

The MCS method was also leveraged to estimate the network risk based on the same number of unique system states explored in the TMCMC method (i.e., 44,124 samples). The estimated network risk with the MCS method was 32.57\%. The close agreement between the TMCMC and MCS results suggests that the system risk is not heavily influenced by gray swan events. This may be attributed to the assumed range of reliability indices (zero to three) and/or the random assignment of reliability indices to bridge links.

Beyond risk estimation, the proposed TMCMC method provides valuable insights through the samples generated in the final stage. The asset-level importance measure as given in \mbox{Eq. \eqref{eq:asset_impo}} was calculated for each of the 1,938 vulnerable bridge links. As previously discussed, this measure is a risk-informed importance indicator, considering both the asset failure probability and failure consequence. A higher importance measure indicates that an asset may have a high failure probability, a significant impact on network capacity, or a combination of both. 

\begin{figure}
    \centering
    \includegraphics[scale=1, trim={0cm 1.5cm 0 1cm}, clip]{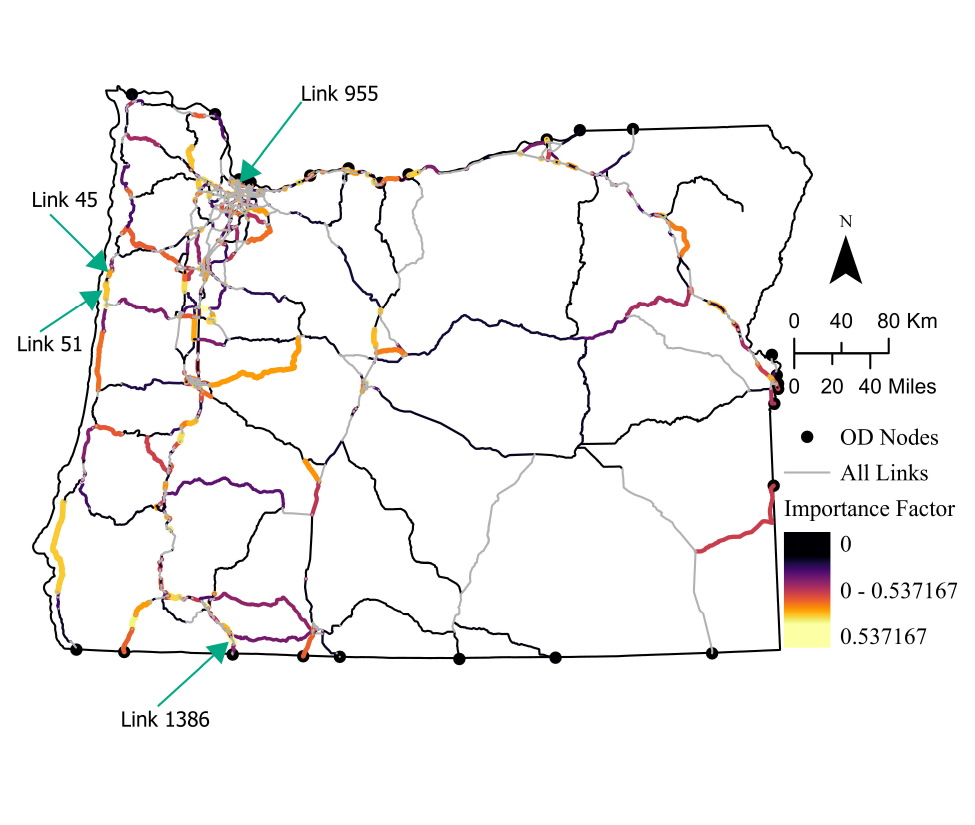}
    \caption{Heatmap of bridge link importance measures.}
    \label{fig:importance_map}
\end{figure}

For instance, consider Link 1386 and Link 955, segments of I5 and OR99E, respectively, shown in Fig. \ref{fig:importance_map}. Link 1386 is located near Oregon's southern border with fewer alternative routes, whereas Link 955 is in the northern part of Portland with multiple alternative paths. Their failure probabilities were 0.469 (for Link 1386) and 0.480 (for Link 955). Based on the TMCMC samples, the importance measure of \mbox{Link 1386} was 0.524 whereas that of Link 955 was 0.430. This indicates that Link 1386 is more important for preserving network capacity due to its lack of alternative links despite its lower probability of failure compared to Link 955.

For contrast, two consecutive segments of US 101, Link 51 and 45, are also shown in Fig. \ref{fig:importance_map}. Their failure probabilities were 0.477 (for Link 51) and 0.378 (for Link 45), respectively. Link 51 connects Newport to Depoe Bay while Link 45 connects Depoe Bay to Kernville. Considering all OD pairs, the two links should have the same impact on network capacity because the flow on one link has to pass the other link too. The TMCMC samples indicate that their importance measures are 0.464 (for Link 51) and 0.423 (for Link 45), respectively. This illustrates the case when a higher failure probability can increase the importance of an asset. Besides the four links described previously,  Fig. \ref{fig:importance_map} shows a heat-map of risk-informed importance measures associated with all links. A bridge link with a higher importance measure may indicate prioritized interventions in order to effectively mitigate indirect risk. 

\section{Conclusions}

In this paper, a novel approach to risk assessment of large-scale transportation networks has been proposed. The risk assessment focuses on variations in system performance under adverse events. Network capacity was analyzed as an example of the system-level performance indicators. The effects of a generic adverse event were modeled by failure probabilities of vulnerable assets under the event. By reformulating the system risk as the evidence term in Bayesian updating, a new method has been developed for risk assessment based on the TMCMC method, a sequential sampling technique especially suitable for large-scale systems. In order to demonstrate the advantages of the new approach, two types of analytical examples with closed-form ground-truth risks were devised to compare it with existing methods for indirect risk assessment. The approach was also applied to evaluate the indirect risk of the Oregon highway network under a hypothetical set of asset failure probabilities. From the results of the analyses, the following conclusions can be drawn: 
\begin{itemize}
    \item Risk of transportation infrastructure in terms of system performance can be reformulated as a Bayesian updating problem, where the posterior likelihood of a system state (i.e., a pattern of asset failures) reflects the performance risk associated with that system state. This reformulation allows for the TMCMC method to efficiently and effectively estimate the risk of large-scale transportation systems. The posterior failure probability of each asset also indicates the importance of the asset. This importance measure offers valuable insights for transportation asset management by identifying critical assets and enabling more effective resource allocation for maintenance and resilience planning.
    \item A series of analytical examples (Case I) were devised by assuming a performance indicator that aggregates asset-level failure consequences via summation. This allows for the precise calculation of system risk. These analytical problems were utilized to test the effectiveness of the new approach with respect to a growing number of assets within a system. The effectiveness was compared with other simulation-based and non-simulation-based methods for risk assessment. It was found that both the proposed and the conventional MC methods can perform well given this hypothetical performance indicator, regardless of asset numbers. However, non-simulation-based methods, which rely on heuristically finding and analyzing a critical subset of system states, become infeasible even for a moderately sized network (e.g., a system with 30 assets). 
    \item To further differentiate the proposed and the conventional MC methods, another set of analytical examples (Case II) was formulated for a hypothetical system involving 30 and 50 assets. In these examples, the performance indicator was designed such that the system risk can still be determined analytically, yet it is controlled by a small set of low likelihood, high consequence events. Results from these examples indicate that due to the existence of gray swan events, the conventional MC method may dramatically underestimate the system risk, and/or the estimated risk may involve an impractical amount of variation. By contrast, the proposed method could still accurately estimate the system risk.
    \item To test the scalability of the proposed approach, the risk of the Oregon highway network was assessed in terms of drop in network capacity. The studied network included 6,437 nodes, 10,637 links, and 1,938 vulnerable bridge links. It was shown that the new approach can still perform in a network of this realistic scale. In addition to the overall risk, the new approach can also be used to derive risk-informed importance measures for assets. Results from the case study indicated that an asset with a high importance measure may either have a high failure probability, or high influence on network capacity, or both. This information can provide valuable insight into performance-based and risk-informed transportation asset management.
    \item The risk assessment method proposed herein was based on the assumptions that the asset failures are independent. However, failure of an asset may implicate higher failure probabilities of nearby assets. Future work in this regard is warranted. 
    \item The proposed approach may potentially be applied beyond transportation networks to any infrastructure systems with interconnected assets, such as water distribution networks and electrical power grids, provided that the system performance is appropriately redefined.
\end{itemize}

\section{Data Availability Statement}

Some or all data, models, or code that support the findings of this study are available from the corresponding author on reasonable request.

\section{Acknowledgments}

The authors are grateful for the financial support received from Portland State University. The opinions and conclusions presented in this paper are those of the authors and do not necessarily reflect the views of the sponsoring organization.

%
%


%
%
\pagebreak
\bibliography{ascexmpl-new}

\begin{thebibliography}{}

\bibitem[\protect\citeauthoryear{}{AASHTO}{2021}]{AASHTOWare}
AASHTO (2021).
\newblock ``{AASHTOWare Bridge Management}.''\ {\em American Association of State Highway and Transportation Officials (AASHTO)}, $<$https://www.aashtowarebridge.com$>$.
\newblock Accessed: 2021-11-20.

\bibitem[\protect\citeauthoryear{}{Ahuja et~al.\@}{1988}]{ahuja1988}
Ahuja, R.~K., Magnanti, T.~L., and Orlin, J.~B. (1988).
\newblock {\em Network Flows}.
\newblock Alfred P. Sloan School of Management, Massachusetts Institute of Technology, Cambridge, MA.

\bibitem[\protect\citeauthoryear{}{Allen et~al.\@}{2022}]{allenSensitivity2022}
Allen, E., Chamorro, A., Poulos, A., Castro, S., {Carlos de la Llera}, J., and Echaveguren, T. (2022).
\newblock ``Sensitivity analysis and uncertainty quantification of a seismic risk model for road networks.''\ {\em Computer-aided Civil and Infrastructure Engineering}, 37(4), 516--530.

\bibitem[\protect\citeauthoryear{}{Au and Beck}{2003}]{auImportant2003}
Au, S.-K. and Beck, J.~L. (2003).
\newblock ``Important sampling in high dimensions.''\ {\em Structural Safety}, 25(2), 139--163.

\bibitem[\protect\citeauthoryear{}{Beck and Au}{2002}]{beckBayesian2002}
Beck, J.~L. and Au, S.~K. (2002).
\newblock ``Bayesian updating of structural models and reliability using {{Markov}} chain {{Monte Carlo}} simulation.''\ {\em ASCE Journal of Engineering Mechanics}, 128(4), 380--391.

\bibitem[\protect\citeauthoryear{}{Betz et~al.\@}{2016}]{betzTransitional2016}
Betz, W., Papaioannou, I., and Straub, D. (2016).
\newblock ``Transitional {{Markov Chain Monte Carlo}}: {{Observations}} and {{Improvements}}.''\ {\em ASCE Journal of Engineering Mechanics}, 142(5), 04016016.

\bibitem[\protect\citeauthoryear{}{Bocchini and Frangopol}{2011a}]{bocchiniGeneralized2011}
Bocchini, P. and Frangopol, D.~M. (2011a).
\newblock ``Generalized bridge network performance analysis with correlation and time-variant reliability.''\ {\em Structural Safety}, 33(2), 155--164.

\bibitem[\protect\citeauthoryear{}{Bocchini and Frangopol}{2011b}]{bocchiniStochastic2011}
Bocchini, P. and Frangopol, D.~M. (2011b).
\newblock ``A stochastic computational framework for the joint transportation network fragility analysis and traffic flow distribution under extreme events.''\ {\em Probabilistic Engineering Mechanics}, 26(2), 182--193.

\bibitem[\protect\citeauthoryear{}{Bocchini and Frangopol}{2013}]{bocchiniConnectivityBased2013}
Bocchini, P. and Frangopol, D.~M. (2013).
\newblock ``Connectivity-{{Based Optimal Scheduling}} for {{Maintenance}} of {{Bridge Networks}}.''\ {\em ASCE Journal of Engineering Mechanics}, 139(6), 760--769.

\bibitem[\protect\citeauthoryear{}{Boeing}{2017}]{boeingOSMnx2017}
Boeing, G. (2017).
\newblock ``{{OSMnx}}: {{New}} methods for acquiring, constructing, analyzing, and visualizing complex street networks.''\ {\em Computers, Environment and Urban Systems}, 65, 126--139.

\bibitem[\protect\citeauthoryear{}{Brooks et~al.\@}{2011}]{brooksHandbook2011}
S. Brooks, A. Gelman, G. Jones, and X. Meng, eds. (2011).
\newblock {\em Handbook of {{Markov Chain Monte Carlo}}}.
\newblock CRC Press, Boca Raton, FL, 1st edition.

\bibitem[\protect\citeauthoryear{}{Capacci et~al.\@}{2023}]{capacciProbabilistic2023}
Capacci, L., Biondini, F., and Kiremidjian, A.~S. (2023).
\newblock ``Probabilistic resilience assessment of aging bridge networks based on damage disaggregation and stationary proposal importance sampling.''\ {\em Life-Cycle of Structures and Infrastructure Systems}, F. Biondini and D.~M. Frangopol, eds., CRC Press,  914--922.

\bibitem[\protect\citeauthoryear{}{C{\'e}rou et~al.\@}{2012}]{cerouSequential2012}
C{\'e}rou, F., Del~Moral, P., Furon, T., and Guyader, A. (2012).
\newblock ``Sequential {{Monte Carlo}} for rare event estimation.''\ {\em Statistics and Computing}, 22(3), 795--808.

\bibitem[\protect\citeauthoryear{}{Chan et~al.\@}{2024}]{chan2024adaptive}
Chan, J., Paredes, R., Papaioannou, I., and Duenas-Osorio, L. (2024).
\newblock ``Adaptive monte carlo methods for estimating rare events in power grids.''\ {\em TechRxiv}\ https://doi.org/10.36227/techrxiv.170654653.30299222/v1.

\bibitem[\protect\citeauthoryear{}{Chang}{2010}]{changtransportation2010}
Chang, L. (2010).
\newblock ``Transportation {{System Modeling}} and {{Applications}} in {{Earthquake Engineering}}.''\ Ph.D. thesis, Dept. of Civil and Environmental Engineering, Graduate College of the Univ. of Illinois at Urbana-Champaign, Champaign, IL, USA.

\bibitem[\protect\citeauthoryear{}{Chang et~al.\@}{2012}]{changBridge2012_}
Chang, L., Peng, F., Ouyang, Y., Elnashai, A.~S., and Spencer, B.~F. (2012).
\newblock ``Bridge {{Seismic Retrofit Program Planning}} to {{Maximize Postearthquake Transportation Network Capacity}}.''\ {\em ASCE Journal of Infrastructure Systems}, 18(2), 75--88.

\bibitem[\protect\citeauthoryear{}{Chang et~al.\@}{2000}]{changProbabilisticEarthquakeScenarios2000}
Chang, S.~E., Shinozuka, M., and Moore, J.~E. (2000).
\newblock ``Probabilistic earthquake scenarios: Extending risk analysis methodologies to spatially distributed systems.''\ {\em Earthquake Spectra}, 16(3), 557--572.

\bibitem[\protect\citeauthoryear{}{Ching and Chen}{2007}]{chingTransitional2007}
Ching, J. and Chen, Y.-C. (2007).
\newblock ``Transitional {{Markov Chain Monte Carlo Method}} for {{Bayesian Model Updating}}, {{Model Class Selection}}, and {{Model Averaging}}.''\ {\em ASCE Journal of Engineering Mechanics}, 133(7), 816--832.

\bibitem[\protect\citeauthoryear{}{Der~Kiureghian}{2022}]{derkiureghianStructural2022}
Der~Kiureghian, A. (2022).
\newblock {\em Structural and {{System Reliability}}}.
\newblock Cambridge University Press, Cambridge, UK.

\bibitem[\protect\citeauthoryear{}{{DHS}}{2013}]{DHS2013_infra_plan}
{DHS} (2013).
\newblock {\em NIPP 2013: Partnering for Critical Infrastructure Security and Resilience}.
\newblock Department of Homeland Security (DHS), Washington, DC.

\bibitem[\protect\citeauthoryear{}{Ditlevsen and Madsen}{1996}]{ditlevsenStructural1996}
Ditlevsen, O. and Madsen, H.~O. (1996).
\newblock {\em Structural Reliability Methods}.
\newblock Wiley, Chichester, NY.

\bibitem[\protect\citeauthoryear{}{Dong et~al.\@}{2020}]{dongMeasuring2020}
Dong, S., Mostafizi, A., Wang, H., Gao, J., and Li, X. (2020).
\newblock ``Measuring the {{Topological Robustness}} of {{Transportation Networks}} to {{Disaster-Induced Failures}}: {{A Percolation Approach}}.''\ {\em ASCE Journal of Infrastructure Systems}, 26(2), 04020009.

\bibitem[\protect\citeauthoryear{}{Echard et~al.\@}{2013}]{echardCombined2013}
Echard, B., Gayton, N., Lemaire, M., and Relun, N. (2013).
\newblock ``A combined {{Importance Sampling}} and {{Kriging}} reliability method for small failure probabilities with time-demanding numerical models.''\ {\em Reliability Engineering \& System Safety}, 111, 232--240.

\bibitem[\protect\citeauthoryear{}{FHWA}{2015}]{fhwaSynchronizing2015}
FHWA (2015).
\newblock ``Synchronizing environmental reviews for transportation and other infrastructure projects: 2015 red book.''\ {\em Report No. FHWA-HEP-15-047}, Federal Highway Administration (FHWA), Washington, DC.

\bibitem[\protect\citeauthoryear{}{Ghosh et~al.\@}{2014}]{ghoshSeismic2014}
Ghosh, J., Rokneddin, K., Padgett, J.~E., and {Duenas-Osorio}, L. (2014).
\newblock ``Seismic {{Reliability}}. {{Assessment}} of {{Aging Highway Bridge Networks}} with {{Field Instrumentation Data}} and {{Correlated Failures}}, {{I}}: {{{\emph{Methodology}}}}.''\ {\em Earthquake Spectra}, 30(2), 795--817.

\bibitem[\protect\citeauthoryear{}{Ghosn et~al.\@}{2016}]{ghosnPerformance2016}
Ghosn, M., {Due{\~n}as-Osorio}, L., Frangopol, D.~M., McAllister, T.~P., Bocchini, P., Manuel, L., Ellingwood, B.~R., Arangio, S., Bontempi, F., Shah, M., Akiyama, M., Biondini, F., Hernandez, S., and Tsiatas, G. (2016).
\newblock ``Performance {{Indicators}} for {{Structural Systems}} and {{Infrastructure Networks}}.''\ {\em ASCE Journal of Structural Engineering}, 142(9), F4016003.

\bibitem[\protect\citeauthoryear{}{Guikema and Gardoni}{2009}]{guikemaReliabilityEstimationNetworks2009}
Guikema, S. and Gardoni, P. (2009).
\newblock ``Reliability {{Estimation}} for {{Networks}} of {{Reinforced Concrete Bridges}}.''\ {\em Journal of Infrastructure Systems}, 15(2), 61--69.

\bibitem[\protect\citeauthoryear{}{Haario et~al.\@}{2006}]{haarioDRAM2006}
Haario, H., Laine, M., Mira, A., and Saksman, E. (2006).
\newblock ``{{DRAM}}: {{Efficient}} adaptive {{MCMC}}.''\ {\em Statistics and Computing}, 16(4), 339--354.

\bibitem[\protect\citeauthoryear{}{Haario et~al.\@}{2001}]{haarioAdaptive2001}
Haario, H., Saksman, E., and Tamminen, J. (2001).
\newblock ``An {{Adaptive Metropolis Algorithm}}.''\ {\em Bernoulli}, 7(2), 223--242.

\bibitem[\protect\citeauthoryear{}{Hagberg et~al.\@}{2008}]{hagbergExploring2008}
Hagberg, A.~A., Schult, D.~A., and Swart, P.~J. (2008).
\newblock ``Exploring network structure, dynamics, and function using networkx.''\ {\em Proc., 7th Python in Science Conf.}, G. Varoquaux, T. Vaught, and J. Millman, eds., Austin, TX.
\newblock SciPy.

\bibitem[\protect\citeauthoryear{}{Hastings}{1970}]{hastingsMonte1970}
Hastings, W.~K. (1970).
\newblock ``Monte {{Carlo Sampling Methods Using Markov Chains}} and {{Their Applications}}.''\ {\em Biometrika}, 57(1), 97--109.

\bibitem[\protect\citeauthoryear{}{Huang et~al.\@}{2019}]{huangStateoftheart2019}
Huang, Y., Shao, C., Wu, B., Beck, J.~L., and Li, H. (2019).
\newblock ``State-of-the-art review on {{Bayesian}} inference in structural system identification and damage assessment.''\ {\em Advances in Structural Engineering}, 22(6), 1329--1351.

\bibitem[\protect\citeauthoryear{}{Ishibashi et~al.\@}{2020}]{ishibashiFramework2020}
Ishibashi, H., Akiyama, M., Frangopol, D.~M., Koshimura, S., Kojima, T., and Nanami, K. (2020).
\newblock ``Framework for estimating the risk and resilience of road networks with bridges and embankments under both seismic and tsunami hazards.''\ {\em Structure and Infrastructure Engineering}.

\bibitem[\protect\citeauthoryear{}{Jafino et~al.\@}{2020}]{jafinotransport2020}
Jafino, B.~A., Kwakkel, J., and Verbraeck, A. (2020).
\newblock ``Transport network criticality metrics: A comparative analysis and a guideline for selection.''\ {\em Transport Reviews}, 40(2), 241--264.

\bibitem[\protect\citeauthoryear{}{Kang et~al.\@}{2012}]{kangFurtherDevelopmentMatrixbased2012}
Kang, W.-H., Lee, Y.-J., Song, J., and Gencturk, B. (2012).
\newblock ``Further development of matrix-based system reliability method and applications to structural systems.''\ {\em Structure and Infrastructure Engineering}, 8(5), 441--457.

\bibitem[\protect\citeauthoryear{}{Kang et~al.\@}{2008}]{kangMatrixbasedSystemReliability2008}
Kang, W.-H., Song, J., and Gardoni, P. (2008).
\newblock ``Matrix-based system reliability method and applications to bridge networks.''\ {\em Reliability Engineering \& System Safety}, 93(11), 1584--1593.

\bibitem[\protect\citeauthoryear{}{Kim et~al.\@}{2013}]{kimSystemReliabilityAnalysis2013}
Kim, D.-S., Ok, S.-Y., Song, J., and Koh, H.-M. (2013).
\newblock ``System reliability analysis using dominant failure modes identified by selective searching technique.''\ {\em Reliability Engineering \& System Safety}, 119, 316--331.

\bibitem[\protect\citeauthoryear{}{Kim et~al.\@}{2012}]{kimAssessmentSeismicRisk2012}
Kim, Y., Kang, W.-H., and Song, J. (2012).
\newblock ``Assessment of seismic risk and importance measures of interdependent networks using a non simulation-based method.''\ {\em Journal of Earthquake Engineering}, 16(6), 777--794.

\bibitem[\protect\citeauthoryear{}{Kiremidjian et~al.\@}{2007a}]{kiremidjianSeismic2007}
Kiremidjian, A., Moore, J., Fan, Y.~Y., Yazlali, O., Basoz, N., and Williams, M. (2007a).
\newblock ``Seismic {{Risk Assessment}} of {{Transportation Network Systems}}.''\ {\em Journal of Earthquake Engineering}, 11(3), 371--382.

\bibitem[\protect\citeauthoryear{}{Kiremidjian et~al.\@}{2007b}]{kiremidjianIssues2007}
Kiremidjian, A.~S., Stergiou, E., and Lee, R. (2007b).
\newblock ``Issues in {{Seismic Risk Assessment}} of {{Transportation Networks}}.''\ {\em Earthquake {{Geotechnical Engineering}}}, A. Ansal and K.~D. Pitilakis, eds., Vol.~6, Springer, Dordrecht, Netherlands,  461--480.

\bibitem[\protect\citeauthoryear{}{Lam et~al.\@}{2020}]{LamImpact2020}
Lam, J.~C., Hackl, J., Heitzler, M., Adey, B.~T., and Hurni, L. (2020).
\newblock ``Impact assessment of extreme hydrometeorological hazard events on road networks.''\ {\em Journal of Infrastructure Systems}, 26(2), 04020005.

\bibitem[\protect\citeauthoryear{}{Lu et~al.\@}{2023}]{luSimulation2023}
Lu, T., Capacci, L., Anghileri, M., Bianchi, S., Luo, D., and Biondini, F. (2023).
\newblock ``Simulation-based seismic risk and robustness assessment of ageing bridge networks.''\ {\em International Journal of Critical Infrastructures}, 19(6), 578--602.

\bibitem[\protect\citeauthoryear{}{Lye et~al.\@}{2021}]{lyeSampling2021}
Lye, A., Cicirello, A., and Patelli, E. (2021).
\newblock ``Sampling methods for solving {{Bayesian}} model updating problems: {{A}} tutorial.''\ {\em Mechanical Systems and Signal Processing}, 159, 107760.

\bibitem[\protect\citeauthoryear{}{Messore et~al.\@}{2021}]{messoreLifecycleCostbasedRisk2021}
Messore, M.~M., Capacci, L., and Biondini, F. (2021).
\newblock ``Life-cycle cost-based risk assessment of aging bridge networks.''\ {\em Structure and Infrastructure Engineering}, 17(4), 515--533.

\bibitem[\protect\citeauthoryear{}{Morlok and Chang}{2004}]{morlokMeasuring2004}
Morlok, E.~K. and Chang, D.~J. (2004).
\newblock ``Measuring capacity flexibility of a transportation system.''\ {\em Transportation Research Part A: Policy and Practice}, 38(6), 405--420.

\bibitem[\protect\citeauthoryear{}{Morrison et~al.\@}{2022}]{morrisonExploring2022}
Morrison, D., Bedinger, M., Beevers, L., and McClymont, K. (2022).
\newblock ``Exploring the raison d'etre behind metric selection in network analysis: A systematic review.''\ {\em Applied Network Science}, 7(1), 1--25.

\bibitem[\protect\citeauthoryear{}{Newman}{2018}]{newmanNetworks2018}
Newman, M. (2018).
\newblock {\em Networks}.
\newblock Oxford University Press, Oxford, NY, 2nd edition.

\bibitem[\protect\citeauthoryear{}{Nicholson et~al.\@}{2016}]{nicholsonFlowbased2016}
Nicholson, C.~D., Barker, K., and {Ramirez-Marquez}, J.~E. (2016).
\newblock ``Flow-based vulnerability measures for network component importance: {{Experimentation}} with preparedness planning.''\ {\em Reliability Engineering \& System Safety}, 145, 62--73.

\bibitem[\protect\citeauthoryear{}{Nojima}{1998}]{nojimaPrioritization1998}
Nojima, N. (1998).
\newblock ``Prioritization in upgrading seismic performance of road network based on system reliability analysis.''\ {\em Proc., 3rd China-Japan-US Trilateral Symp. on Lifeline Earthquake Engineering}, Reston, VA, American Society of Civil Engineers,  323--330.

\bibitem[\protect\citeauthoryear{}{Olivier et~al.\@}{2020}]{olivierUQpy2020}
Olivier, A., Giovanis, D.~G., Aakash, B.~S., Chauhan, M., Vandanapu, L., and Shields, M.~D. (2020).
\newblock ``Uqpy: A general purpose python package and development environment for uncertainty quantification.''\ {\em Journal of Computational Science}.

\bibitem[\protect\citeauthoryear{}{Proctor and Varma}{2012}]{proctorRiskBased2012}
Proctor, G.~D. and Varma, S. (2012).
\newblock ``Risk-{{Based Transportation Asset Management}}.''\ {\em Report No. FHWA-HIF-12-036}, Washington, DC: Federal Highway Administration (FHWA).

\bibitem[\protect\citeauthoryear{}{Ren and Orkoulas}{2007}]{renParallel2007}
Ren, R. and Orkoulas, G. (2007).
\newblock ``Parallel {{Markov}} chain {{Monte Carlo}} simulations.''\ {\em Journal of Chemical Physics}, 126(21), 211102.

\bibitem[\protect\citeauthoryear{}{Robert and Casella}{2005}]{robertMonteCarloStatistical2005}
Robert, C. and Casella, G. (2005).
\newblock {\em Monte {{Carlo Statistical Methods}}}.
\newblock Springer, New York, 2nd edition.

\bibitem[\protect\citeauthoryear{}{Robert et~al.\@}{2018}]{robertAccelerating2018}
Robert, C.~P., Elvira, V., Tawn, N., and Wu, C. (2018).
\newblock ``Accelerating {{MCMC}} algorithms.''\ {\em WIREs Computational Statistics}, 10(5), e1435.

\bibitem[\protect\citeauthoryear{}{Roberts and Rosenthal}{2009}]{robertsExamples2009}
Roberts, G.~O. and Rosenthal, J.~S. (2009).
\newblock ``Examples of {{Adaptive MCMC}}.''\ {\em Journal of Computational and Graphical Statistics}, 18(2), 349--367.

\bibitem[\protect\citeauthoryear{}{Rokneddin et~al.\@}{2014}]{rokneddinSeismic2014}
Rokneddin, K., Ghosh, J., {Due{\~n}as-Osorio}, L., and Padgett, J.~E. (2014).
\newblock ``Seismic {{Reliability Assessment}} of {{Aging Highway Bridge Networks}} with {{Field Instrumentation Data}} and {{Correlated Failures}}, {{II}}: {{Application}}.''\ {\em Earthquake Spectra}, 30(2), 819--843.

\bibitem[\protect\citeauthoryear{}{Rubin et~al.\@}{2015}]{rubinBayesian2015}
Rubin, J.~B., Carlin, H.~S., Stern, D.~B., Dunson, A.~V., Rubin, D.~B., and Gelman, A. (2015).
\newblock {\em Bayesian Data Analysis}.
\newblock Chapman and Hall/CRC, NY, 3 edition.

\bibitem[\protect\citeauthoryear{}{Saydam et~al.\@}{2013}]{saydamTimedependent2013}
Saydam, D., Bocchini, P., and Frangopol, D.~M. (2013).
\newblock ``Time-dependent risk associated with deterioration of highway bridge networks.''\ {\em Engineering Structures}, 54, 221--233.

\bibitem[\protect\citeauthoryear{}{Serdar et~al.\@}{2022}]{serdarUrban2022}
Serdar, M.~Z., Ko{\c c}, M., and {Al-Ghamdi}, S.~G. (2022).
\newblock ``Urban {{Transportation Networks Resilience}}: {{Indicators}}, {{Disturbances}}, and {{Assessment Methods}}.''\ {\em Sustainable Cities and Society}, 76, 103452.

\bibitem[\protect\citeauthoryear{}{Shiraki et~al.\@}{2007}]{shirakiSystem2007}
Shiraki, N., Shinozuka, M., Moore, J.~E., Chang, S.~E., Kameda, H., and Tanaka, S. (2007).
\newblock ``System risk curves: Probabilistic performance scenarios for highway networks subject to earthquake damage.''\ {\em ASCE Journal of Infrastructure Systems}, 13(1), 43--54.

\bibitem[\protect\citeauthoryear{}{Simoen et~al.\@}{2015}]{simoenDealing2015}
Simoen, E., De~Roeck, G., and Lombaert, G. (2015).
\newblock ``Dealing with uncertainty in model updating for damage assessment: {{A}} review.''\ {\em Mechanical Systems and Signal Processing},  123--149.

\bibitem[\protect\citeauthoryear{}{Soleimani et~al.\@}{2021}]{soleimaniMultihazard2021}
Soleimani, N., Davidson, R.~A., Davis, C., O'Rourke, T.~D., and Nozick, L.~K. (2021).
\newblock ``Multihazard {{Scenarios}} for {{Regional Seismic Risk Assessment}} of {{Spatially Distributed Infrastructure}}.''\ {\em Journal of Infrastructure Systems}, 27(1), 04021001.

\bibitem[\protect\citeauthoryear{}{Song and Kawai}{2023}]{songMonte2023}
Song, C. and Kawai, R. (2023).
\newblock ``Monte {{Carlo}} and variance reduction methods for structural reliability analysis: {{A}} comprehensive review.''\ {\em Probabilistic Engineering Mechanics}, 73, 103479.

\bibitem[\protect\citeauthoryear{}{Song and Kang}{2009}]{songSystemReliabilitySensitivity2009}
Song, J. and Kang, W.-H. (2009).
\newblock ``System reliability and sensitivity under statistical dependence by matrix-based system reliability method.''\ {\em Structural Safety}, 31(2), 148--156\ Risk Acceptance and Risk Communication.

\bibitem[\protect\citeauthoryear{}{Song et~al.\@}{2020}]{songAccounting2020}
Song, M., Behmanesh, I., Moaveni, B., and Papadimitriou, C. (2020).
\newblock ``Accounting for {{Modeling Errors}} and {{Inherent Structural Variability}} through a {{Hierarchical Bayesian Model Updating Approach}}: {{An Overview}}.''\ {\em Sensors}, 20(14), 3874.

\bibitem[\protect\citeauthoryear{}{Song et~al.\@}{2009}]{songSubset2009}
Song, S., Lu, Z., and Qiao, H. (2009).
\newblock ``Subset simulation for structural reliability sensitivity analysis.''\ {\em Reliability Engineering \& System Safety}, 94(2), 658--665.

\bibitem[\protect\citeauthoryear{}{Temme}{2024}]{temme2024}
Temme, N.~M. (2024).
\newblock {\em Error Functions, Dawson’s and Fresnel Integrals}.
\newblock The National Institute of Standards and Technology (NIST).
\newblock Online.

\bibitem[\protect\citeauthoryear{}{Vishnu et~al.\@}{2023}]{vishnuRoad2023}
Vishnu, N., Kameshwar, S., and Padgett, J.~E. (2023).
\newblock ``Road transportation network hazard sustainability and resilience: Correlations and comparisons.''\ {\em Structure and Infrastructure Engineering}, 19(3), 345--365.

\bibitem[\protect\citeauthoryear{}{Western et~al.\@}{2016}]{westernj}
Western, J., Bye, P., Valeo, M., Thompson, P., and Frazier, E. (2016).
\newblock ``Final report: Assessing risk for bridge management.''\ {\em Report No. NCHRP 20-07/Task 378}, National Cooperative Highway Research Program (NCHRP), Washington, DC.

\bibitem[\protect\citeauthoryear{}{Yang}{2022}]{yangDeep2022}
Yang, D.~Y. (2022).
\newblock ``Deep reinforcement learning-enabled bridge management considering asset and network risks.''\ {\em ASCE Journal of Infrastructure Systems}, 28(3), 04022023.

\bibitem[\protect\citeauthoryear{}{Yang and Frangopol}{2018}]{yangRiskInformed2018}
Yang, D.~Y. and Frangopol, D.~M. (2018).
\newblock ``Risk-informed bridge ranking at project and network levels.''\ {\em ASCE Journal of Infrastructure Systems}, 24(3), 04018018.

\bibitem[\protect\citeauthoryear{}{Yang and Frangopol}{2020a}]{yangLifecycle2020}
Yang, D.~Y. and Frangopol, D.~M. (2020a).
\newblock ``Life-cycle management of deteriorating bridge networks with network-level risk bounds and system reliability analysis.''\ {\em Structural Safety}, 83, 101911.

\bibitem[\protect\citeauthoryear{}{Yang and Frangopol}{2020b}]{yangRiskbased2020}
Yang, D.~Y. and Frangopol, D.~M. (2020b).
\newblock ``Risk-based portfolio management of civil infrastructure assets under deep uncertainties associated with climate change: A robust optimisation approach.''\ {\em Structure and Infrastructure Engineering}, 16(4), 531--546.

\bibitem[\protect\citeauthoryear{}{Yang and Frangopol}{2022}]{yangRiskbased2022}
Yang, D.~Y. and Frangopol, D.~M. (2022).
\newblock ``Risk-based inspection planning of deteriorating structures.''\ {\em Structure and Infrastructure Engineering}, 18(1), 109--128.

\bibitem[\protect\citeauthoryear{}{Zhang and Wang}{2016}]{zhangResiliencebasedRiskMitigation2016}
Zhang, W. and Wang, N. (2016).
\newblock ``Resilience-based risk mitigation for road networks.''\ {\em Structural Safety}, 62, 57--65.

\bibitem[\protect\citeauthoryear{}{Zhou et~al.\@}{2019a}]{zhouconnectivity2019}
Zhou, Y., Wang, J., and Sheu, J.-B. (2019a).
\newblock ``On connectivity of post-earthquake road networks.''\ {\em Transportation Research Part E: Logistics and Transportation Review}, 123, 1--16.

\bibitem[\protect\citeauthoryear{}{Zhou et~al.\@}{2019b}]{zhouResilience2019}
Zhou, Y., Wang, J., and Yang, H. (2019b).
\newblock ``Resilience of {{Transportation Systems}}: {{Concepts}} and {{Comprehensive Review}}.''\ {\em IEEE Transactions on Intelligent Transportation Systems}, 20(12), 4262--4276.

\end{thebibliography}
%
\end{document}